\documentclass[conference]{IEEEtran}
\IEEEoverridecommandlockouts
\usepackage{cite}
\usepackage{amsmath,amssymb,amsfonts}
\usepackage{algorithmic}
\usepackage{graphicx}
\usepackage{textcomp}
\usepackage{xcolor}
\usepackage{algorithm}
\usepackage{makecell}
\usepackage{graphicx}
\usepackage{subfigure}
\usepackage{etoolbox}
\usepackage{multirow}
\usepackage{hyperref}
\usepackage{fancyhdr} 

\hypersetup{hidelinks}

\makeatletter
\patchcmd{\@maketitle}
  {\addvspace{0.5\baselineskip}\egroup}
  {\addvspace{-0.9 \baselineskip}\egroup}
  {}
  {}
\makeatother

\begin{document}
\title{ \vspace{0pt} 
Joint Sensing and Communications for Deep Reinforcement Learning-based Beam Management in 6G
}

\author{\IEEEauthorblockN{Yujie Yao, Hao Zhou, and Melike Erol-Kantarci, \IEEEmembership{Senior Member, IEEE}}
\IEEEauthorblockA{\textit{School of Electrical Engineering and Computer Science} \\
\textit{University of Ottawa},
Emails:\{yyao016,hzhou098, melike.erolkantarci\}@uottawa.ca}} 

\maketitle

\thispagestyle{fancy} %
      \lhead{} 
      \chead{Accepted by 2022 IEEE Globecom conference, \copyright2022 IEEE } 
      \rhead{} 
      \lfoot{} 
      \cfoot{\thepage} 
      \rfoot{} 
      \renewcommand{\headrulewidth}{0pt} 
      \renewcommand{\footrulewidth}{0pt} 
\pagestyle{fancy}

\begin{abstract}
User location is a piece of critical information for network management and control. However, location uncertainty is unavoidable in certain settings leading to localization errors. In this paper, we consider the user location uncertainty in the mmWave networks, and investigate joint vision-aided sensing and communications using deep reinforcement learning-based beam management for future 6G networks. In particular, we first extract pixel characteristic-based features from satellite images to improve localization accuracy. Then we propose a UK-medoids based method for user clustering with location uncertainty, and the clustering results are consequently used for the beam management. Finally, we apply the DRL algorithm for intra-beam radio resource allocation. The simulations first show that our proposed vision-aided method can substantially reduce the localization error. The proposed UK-medoids and DRL based scheme (UKM-DRL) is compared with two other schemes: K-means based clustering and DRL based resource allocation (K-DRL) and UK-means based clustering and DRL based resource allocation (UK-DRL). The proposed method has 17.2\% higher throughput and 7.7\% lower delay than UK-DRL, and more than doubled throughput and 55.8\% lower delay than K-DRL.
\end{abstract}


\section{Introduction}

\begin{figure*}[!t]
\vspace{0pt}
    \centering
    \includegraphics[width=0.98\textwidth]{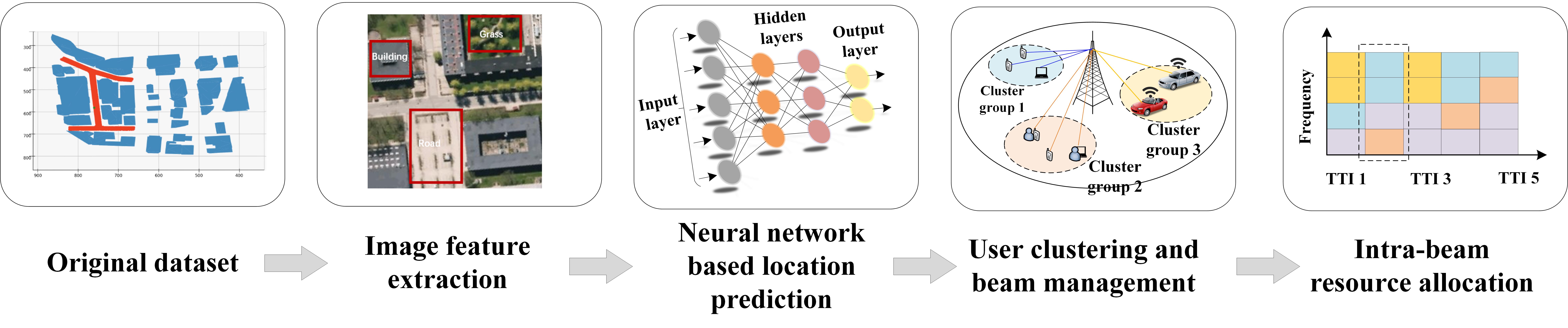}
    \caption{Overall System Architecture
    \label{fig1-1}}
        \vspace{-15pt}
\end{figure*}

Future 6G networks rely on millimeter wave (mmWave) bands to provide rich bandwidth, in which highly directional beams are designed to maintain connections between the base stations (BSs) and user equipment (UEs). However, to apply the mmWave bands in future 6G networks, it requires intelligent beam management schemes to overcome the challenges in a dynamic environment such as inefficient beam sweeping, user mobility and location uncertainty, and so on. In beam management techniques, clustering is generally performed based on the location of users, and each cluster is served by a separate beam \cite{b3}. BSs need to keep track of the user's trajectory and dynamically adjust the direction of beams to serve all UEs. Global Positioning System (GPS) is one well-known technology for outdoor localization. However, the application of GPS is limited to line of sight (LOS) scenarios where GPS becomes inefficient in urban canyons and certain other settings. On the other hand, the received signal-based outdoor localization schemes suffer from propagation loss in mmWave network.

Recently, computer vision is introduced to augment localization in wireless communications, and the visual information can address challenges in the traditional localization approaches \cite{b4}. For example, vision-aided sensing has been successfully applied for robot localization and navigation in \cite{b4-1}. On the other hand, despite the improvements obtained by advanced localization techniques, localization error is still unavoidable in practice, which may be caused by measurement accuracy, system performance fluctuation, and so on. Consequently, the user's location errors may further affect the network performance. For example, if the BS observes distorted locations of users, some users may not be covered by the highly directional beams from BS, which will result in packet loss or longer delay. Therefore, a clustering algorithm that can handle location uncertainty is expected to mitigate the effect of user localization error \cite{B4_3}. Finally, network management becomes more and more complicated with evolving network architecture, and machine learning, especially reinforcement learning, becomes significant solutions for network optimization and control \cite{bb}.  

To this end, we first introduce a vision-aided sensing-based localization method, which utilizes satellite images to predict the user locations. Then we propose a UK-medoids based clustering technique for higher clustering accuracy for uncertain objects. Afterwards, we apply deep reinforcement learning (DRL) algorithm for intra-beam radio resource allocation to satisfy Quality-of-Service (QoS) requirements of users \cite{b7}. In terms of localization performance, the simulations show that our vision-aided method can significantly reduce the localization error from 17.11 meters to 3.62 meters. For the network performance, we compare the proposed UK-medoids and DRL method (UKM-DRL) with the other two baseline algorithms, and our UKM-DRL method achieves higher throughput and lower delay.

The main contributions of this work are two-fold: firstly, we propose a novel vision-aided sensing-based localization method, and secondly, we propose a UKM-DRL algorithm for joint beam management and resource allocation of mmWave networks with user location uncertainty. The remaining of this paper is organized as follows. Section \ref{s2} presents related work, and Section \ref{s3} shows the network system model. Section \ref{s4} introduces the dataset and vision-aided technique. Section \ref{s5} explains the UK-medoids based clustering methods and DRL algorithm. Section \ref{s6} gives simulation results, and Section \ref{s7} concludes this paper.

\section{Related Work}
\label{s2}
User location is one of the most important information for network management\cite{B4_4}. Although the GPS technology has been generally applied, it suffers from degradation caused by none line of sight (NLOS) transmissions in dense urban areas. Recently, vision-aided localization techniques have been proposed as promising alternatives. For example, \cite{b8} proposed an improved localization algorithm based on GPS, inertial navigation system, and visual localization, where the visual information is used to compensate for the instability of GPS. In \cite{b9}, 3D point cloud acquired from depth image is applied to localize and navigate robots, and the authors proposed a fast sampling plane filtering algorithm to reduce the volume of 3D cloud. Meanwhile, \cite{b10} proposed two localization schemes based on deep learning and landmark detection to replace GPS. On the other hand, user clustering is an important part of mmWave network for beamforming. A non-orthogonal multiple access system is considered in \cite{b11}, and the authors optimized the sum rate with a K-means-based algorithm. \cite{B12} proposed a user clustering and power allocation algorithm to maximize the sum capacity, and a hierarchical clustering technique is proposed in \cite{b13} to obtain the optimal number of clusters instead of having a deterministic number. 

Different than the aforementioned works, we first define a computer vision-based localization method, then we propose a UK-medoids based clustering algorithm for beam management that considers the location uncertainty, and a DRL based method for intra-beam resource management. Moreover, we systematically investigate how these technique combinations can improve the mmWave network performance.    

\section{System and Wireless Environment Model}
\label{s3}
\subsection{Overall System Architecture}
The overall system model of this work is illustrated in Fig. \ref{fig1-1}. First, given the original dataset, we extract the image features of each episode in the dataset. Then, these features are used as input for a feed-forward neural network, which predicts the user locations. Next, considering the localization uncertainty, we propose a UK-medoids-based clustering method to form cluster groups for beam management. Finally, we apply DRL for the intra-beam resource allocation of each beam.       

\subsection{Wireless Environment Model}
We consider a 5G-NodeB (gNB) serving a group of users with different QoS requirements. The UE set is denoted by \textbf{\emph{U}} and each UE is represented by $U_i$. 
Based on observed locations, gNB will produce a certain number of UE clusters, and each cluster will be served by one beam. Here the inter-beam interference can be neglected due to the Orthogonal Frequency Division Multiple Access (OFDMA) technique. In each beam, radio resource is allocated to distribute the Resource Block Groups (RBGs) to the attached users.

\begin{figure}
\vspace{-5pt}
    \centering
    \includegraphics[width=0.4\textwidth]{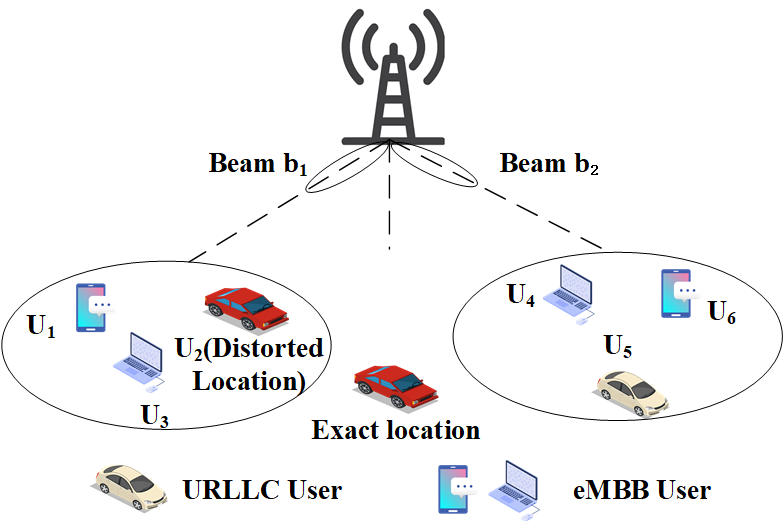}
    \caption{Wireless environment model of mmWave networks
    \label{fig1}}
    \vspace{-15pt}
\end{figure}

As illustrated in Fig. \ref{fig1}, we assume two clusters are formed to cover 6 UEs. We assume the gNB receives distorted location information of the red car $U_2$, and two beams are formed accordingly. Therefore, due to the wrong beam direction, $U_2$ is covered by neither beam $b_1$ nor $b_2$, which results in packet loss and longer delay. This scenario explains how localization error can decrease the network performance of beamforming in mmWave networks. In this work, we mainly focus on the localization and clustering techniques, and a more detailed network and channel model can be found in \cite{b7}.

\section{Satellite Image and Vision-aided Localization Method}
\label{s4}

In this section, we introduce our vision-aided localization method. With the pixel characteristic-based features of satellite images, we aim to improve localization accuracy.

\subsection{Satellite Image Dataset}
The dataset used for localization is based on measurements from a communications system located at the University of Denmark \cite{b14}. In this dataset, one mobile vehicle drives over a region of 2.4 km by 1.25 km and provides measurements, including Signal to Interference and Noise Ratio (SINR), Reference Signal Received Power (RSRP), Reference Signal Received Quality (RSRQ), and Received signal strength indication (RSSI). Each sample contains one satellite image that shows the environment around the vehicle.

\begin{figure}
\vspace{0pt}
    \centering
    \includegraphics[width=0.3\textwidth]{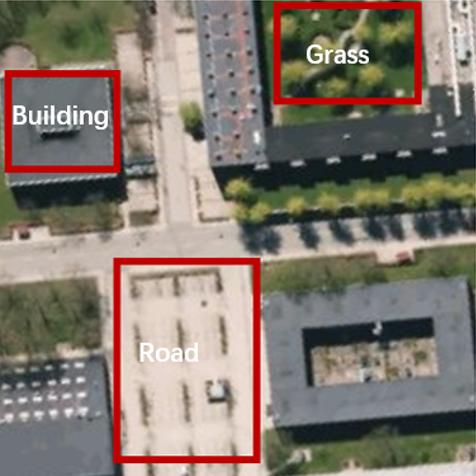}
    \caption{Selected objects in the image of the dataset
    \label{fig2}}
    \vspace{-10pt}
\end{figure}

\begin{table}[t]
\centering
\caption{Mean value of Color Characteristics}
\scalebox{1.1}{
\begin{tabular}{|c|c|c|c|}
    \hline
     &Grass&Building&Road  \\
     \hline
     $|R-G|$&12.14&5.41&15.43\\
     \hline
     $|R-B|$&21.77&8.74&27.31\\
     \hline
     $|G-B|$&31.30&3.39&11.87\\
     \hline
     $|2G-R-B|$&42.13&3.01&4.38\\
     \hline
\end{tabular}}
\label{table1}
\vspace{0pt}
\end{table}

\begin{table}[!t]
\centering
\vspace{-10pt}
\caption{Range of Three Categories}
\scalebox{1.1}{
\begin{tabular}{|c|c|c|c|}
    \hline
     &Grass&Building&Road  \\
     \hline
     $|G-B|$&[0,80]&[0,6]&(6,20]\\
     \hline
     $|2G-R-B|$&[12,85]&[1,8]&[0,12]\\
     \hline
\end{tabular}}
\label{table2}
\vspace{-15pt}
\end{table}

\subsection{Pixel Characteristic-based Feature}
In digital imaging, the domain of the image is partitioned into $X$ rows and $Y$ columns during the sampling process. A pixel represents the area of intersection of one row and one column, which is considered as the smallest element in one picture, then the resolution of the image is $X \times Y$. In RGB color model, each pixel is defined by three values representing red, green, and blue ranging from 0 to 255.

The satellite images used in the localization process are RGB images with the same resolution of $256 \times 256$, and each image is paired with a location sample. Therefore, we count the number of pixels that represent one specific object to compute the proportion of this object in the image. Since there are some common objects in the satellite images such as grass and buildings. We can use this pixel characteristic-based feature to characterize one image. In this work, we selected three objects. As shown in Fig. \ref{fig2}, the three selected objects are grass, gray building, and road. Then we counted the total number of pixels representing three objects respectively as three new features, which will be used to improve the localization accuracy.

To distinguish three objects, we adopted the method from \cite{b16}. 
We selected four color-based characteristics to observe their mean values, where the color characteristics are operations between three basic color values. As shown in Table \ref{table1}, the mean values of $|G-B|$ for the three categories are quite different. Moreover, the mean values of $|2G-R-B|$ are significantly different from the other two categories. After observing the histogram of these two values for three categories, we decide the value range for three categories as shown in Table \ref{table2}.

Finally, we deploy 7 features as input of feed forward neural networks (FFNN), including SINR, RSRP, RSRQ, RSSI, the number of pixels representing grass, gray building, and road. Then the FFNN will be trained to predict the user locations based on the input features \cite{b5}. 

\section{UK-Medoids-based clustering and Deep Reinforcement Learning-based Radio Resource Allocation}
\label{s5}
To handle localization uncertainty, we first deploy the UK-medoids method for clustering, in which each cluster will be covered by one beam \cite{b7}. Then, DRL is applied to distribute physical resource blocks within each beam.
\subsection{UK-medoids}

UK-medoids is an advanced K-medoids based clustering technique \cite{b6}. It selects actual data points as cluster centers instead of computing the mean value of clusters. Meanwhile, the UK-medoids algorithm applies uncertain distance functions that are designed to better estimate the real distance between two uncertain objects. 

It is presumed that there are $M$ positions to form $N$ clusters, and the $j^{th}$ cluster is denoted by $C_j$. The center of each cluster is denoted by $\boldsymbol{c_j}$. For each user position $\boldsymbol{p}$, the uncertainty is represented by a known probability distribution function (PDF) $f(\boldsymbol{p_m})$. We assume there are two uncertain objects $\boldsymbol{p_m}$ and $\boldsymbol{p_n}$, and the uncertain region is $R_m$ and $R_n$, respectively. Then the uncertain distance $\delta(\boldsymbol{p_m},\boldsymbol{p_n})$ is computed by:

\begin{equation}
\resizebox{0.89\hsize}{!}{$
    \delta(\boldsymbol{p_m},\boldsymbol{p_n}) = \int\limits_{\boldsymbol{p_m}\in{R_m}}\int\limits_{\boldsymbol{p_n}\in{R_n}}dist(\boldsymbol{p_m},\boldsymbol{p_n})f(\boldsymbol{p_m})f(\boldsymbol{p_n})d\boldsymbol{p_m}d\boldsymbol{p_n}$},
    \label{eq1}
\end{equation}
where $dist(\boldsymbol{p_m},\boldsymbol{p_n})$ is a distance measure between $\boldsymbol{p_m}$ and $\boldsymbol{p_n}$. The uncertain distance is the expected value of the distance measurement between two uncertain objects. 

Then, the medoid of a cluster is computed as the data point in the cluster with the least average dissimilarity to all other points in the cluster. The update method of a cluster center is:

\begin{equation}
    \boldsymbol{c_j} = \arg \min\limits_{\boldsymbol{p}\in{C_j}}\sum\nolimits_{p^{\prime}\in{C_j}}{\delta(\boldsymbol{p},\boldsymbol{p^{\prime}})}
\label{eq2}
\end{equation}

In this paper, we assume the uncertainty region of all locations is specified as a uniform circle with center $\boldsymbol{p}$ and radius $R$. Then the PDF for a uniformly distributed circle is shown as below in the polar coordinate system:
\begin{equation}
    f(r,\theta)=\frac{r}{\pi{R^2}}
\label{eq3}
\end{equation}
where $r\in{[0,R]}$ and $\pi\in{[0,2\pi]}$.

In this way, equation (\ref{eq1}) can be rewritten as:
\begin{equation}
    \begin{split}
        \delta(\boldsymbol{p_m},\boldsymbol{p_n}) = \int_{0}^{R}{\int_{0}^{2\pi}}\int_{0}^{R}{\int_{0}^{2\pi}}dist(\boldsymbol{p_m},\boldsymbol{p_n})\qquad \quad\\
        f(r,\theta)^2r^2d{r_1}d{\theta_1}d{r_2}d{\theta_2},
    \end{split}
\end{equation}
which is the expansion of the previous equation (\ref{eq1}) from the rectangular coordinate system to the polar coordinate system. In our work, we use Euclidean distance as our distance measurement.

The UK-medoids algorithm is summarized by Algorithm \ref{a1gorithm1}. First, all the uncertain distances between every two objects are computed. Note that these distances only need to be computed once during the whole algorithm, which will reduce the computation complexity. Then, the initial cluster centers are chosen, and it comes to the main loop. The main loop of UK-medoids contains two phases: the first phase is to allocate objects to a cluster, the second phase is to recompute cluster centers according to equation (\ref{eq2}). The loop continues until there is no change in the cluster centers. 

\begin{algorithm}[h]
	\renewcommand{\algorithmicrequire}{\textbf{Input:}}
	\renewcommand{\algorithmicensure}{\textbf{Output:}}
	\caption{UK-Medoids}
	\label{alg1}
	\begin{algorithmic}[1]
		\REQUIRE A set of uncertain location $\boldsymbol{P}$ = $\left\{\boldsymbol{p_1}, ..., \boldsymbol{p_M}\right\}$.
		\ENSURE A set of clusters $\boldsymbol{C}$ = $\left\{\boldsymbol{C_1}, ..., \boldsymbol{C_N}\right\}$.
		\STATE Computing uncertain distance $\delta(\boldsymbol{p_m},\boldsymbol{p_n})$, ${\forall}\boldsymbol{p_m},\boldsymbol{p_n}\in\boldsymbol{P}$
		\STATE Selecting initial cluster centers $\boldsymbol{c}$ = $\left\{\boldsymbol{c_1}, ..., \boldsymbol{c_N}\right\}$.
		
		\WHILE{$\boldsymbol{c}\neq{\boldsymbol{c^{\prime}}}$}
        \STATE $\boldsymbol{c^{\prime}}\gets{\boldsymbol{c}}$
        \STATE $\boldsymbol{c}\gets{\emptyset}$
        \STATE $\boldsymbol{C}$ = $\left\{\boldsymbol{C_1}, ..., \boldsymbol{C_n}\right\}\gets{\left\{\emptyset, ..., \emptyset\right\}}$.
        
		\FOR{Each object $\boldsymbol{p}\in{\boldsymbol{P}}$}
		\STATE Allocating each object to the nearest cluster center according to the uncertain distance.
		\ENDFOR
		\FOR{Each cluster $\boldsymbol{C_j}\in{\boldsymbol{C}}$}
		\STATE Recomputing the cluster center $c_j$.
		\STATE $\boldsymbol{c}\gets{\boldsymbol{c}\cup{c_j}}$
		\ENDFOR

		\ENDWHILE

	\end{algorithmic}  
\label{a1gorithm1}
\end{algorithm}
\subsection{Deep Reinforcement Learning}
UK-Medoids will return a set of clusters, and we assume each cluster will be served by one beam\cite{b7}. Then DRL is performed for intra-beam radio resource allocation, which aims to offer high QoS requirements for both ultra-reliable low latency communications (URLLC) and enhanced mobile broadband (eMBB) users. The Markov decision process of DRL is defined as follows:

1) Actions: In each beam, the action is to allocate $i^{th}$ RBG to $j^{th}$ user.

2) States: The UE's channel quality indicator (CQI) feedback is defined as states, which represents the channel condition between UEs and the BS.

3) Reward: URLLC and eMBB users have different QoS requirements, the reward function must take both requirements into account. URLLC users need minimal latency and high reliability, whereas eMBB users require a higher data rate. As such, the reward function has to consider the requirements of different types of users. 
The reward function is given by:
\begin{equation}
    r=
    \begin{cases}
    sigm({\frac{s_{i,b}}{s^{Tar}}}),& \text{eMBB user}\\
    sigm(\frac{s_{i,b}}{s^{Tar}} \frac{\tau^{Tar}}{\tau^{q}} ),& \text{URLLC user} 
    \end{cases}
    \label{eq12}
\end{equation}
where $s_{i,b}$ is the SINR of the transmission link between $i^{th}$ RBG to a user in the $b^{th}$ beam, $s^{Tar}$ is the target SINR of eMBB users, $\tau^{Tar}$ is the target latency of URLLC users, and $\tau^{q}$ is the queuing delay. Here we use a target SINR to represent the high-reliability requirements of URLLC users. $sigm$ denotes the sigmoid function, which will normalize the reward:
\begin{equation}
    sigm(x)=\frac{1}{1+e^{-x}}
\end{equation} 

Finally, we deploy Long Short-Term Memory (LSTM) network for Q-value prediction in DRL. As a special recurrent neural network, the LSTM network can better capture the long-term data dependency, which makes it an ideal application for complicated wireless environment.

\section{Simulation}
\label{s6}
\subsection{Simulation Settings}
\subsubsection{Localization settings and dataset}
We divide the dataset into 4 regions and each region has its own FFNN to predict the UE locations. The FFNN architecture is $128\times64\times32\times16\times2$. Sigmoid activation functions are employed for hidden layers, whereas linear activation functions are used for output layers.

\subsubsection{Resource Allocation Settings}
We perform the simulation on MATLAB 5G Toolbox. The mmWave network consists of 1 gNB, and the clustering is implemented every 10 TTIs based on the UE trajectories. The traffic of users follows Poisson distribution and the packet size is 32 bytes. Three beams with 60$^\circ$ angle are predefined. The total simulation time is 1400 TTIs, and each TTI is 1/7 ms. The simulation is repeated 5 times, and the confidence interval is 95\%. 

For UEs' positions, the location provided by the dataset is considered as exact location, while the localization results are considered as location with error for comparison. We include four scenarios:
\begin{itemize}
    \item Scenario 1: We applied K-means for clustering and DRL for resource allocation. The scenario is summarized by K-DRL+Location with Error.
    \item Scenario 2: Compared with scenario 1, the only difference is that we use UK-means for clustering. UK-means is another uncertain clustering method that is developed from K-means \cite{b17}. It updates centers as the expected value over PDF and the distance between objects are also the expected distance. The scenario is named by UK-DRL+Location with Error.
    \item Scenario 3: We deploy our proposed UK-medoids based clustering method and DRL for resource allocation. The scenario is named by UKM-DRL+Location with Error. 
    \item Scenario 4: In scenario 4, we use the exact location and K-means for clustering, which is considered as an optimal baseline. This scenario is named by K-DRL+Exact Location.
\end{itemize}
Following we first present the localization and coverage results, then we show the network performance in terms of delay and throughput. 

\subsection{Simulation Results}

\begin{table}[!t]
\vspace{0pt}
\centering
\caption{Measurements of Localization Results}
\scalebox{1.05}{
\begin{tabular}{|c|c|c|c|c|}
\hline
      \multirow{2}*{Results} &\multicolumn{2}{|c|}{Method in \cite{b5}}&\multicolumn{2}{|c|}{Proposed method}\\
    \cline{2-5}
      ~&RMSE&MAE&\makecell{RMSE}&\makecell{MAE}\\
     \hline
     Region 1&11.68&6.03&3.82&1.78\\
     \hline
     Region 2&23.11&10.68&3.52&1.21\\
     \hline
     Region 3&13.68&8.85&1.38&0.97\\
     \hline
     Region 4&16.72&6.12&3.75&1.54\\
     \hline
     Total&17.11&7.83&3.62&1.51\\
     \hline
\end{tabular}}
\label{table5}
\vspace{-10pt}
\end{table}

\begin{figure}[!t]
    \centering
    \subfigure[Coverage rate under varying number of beams]{
    \begin{minipage}[b]{0.4\textwidth}
    \includegraphics[width=1\textwidth]{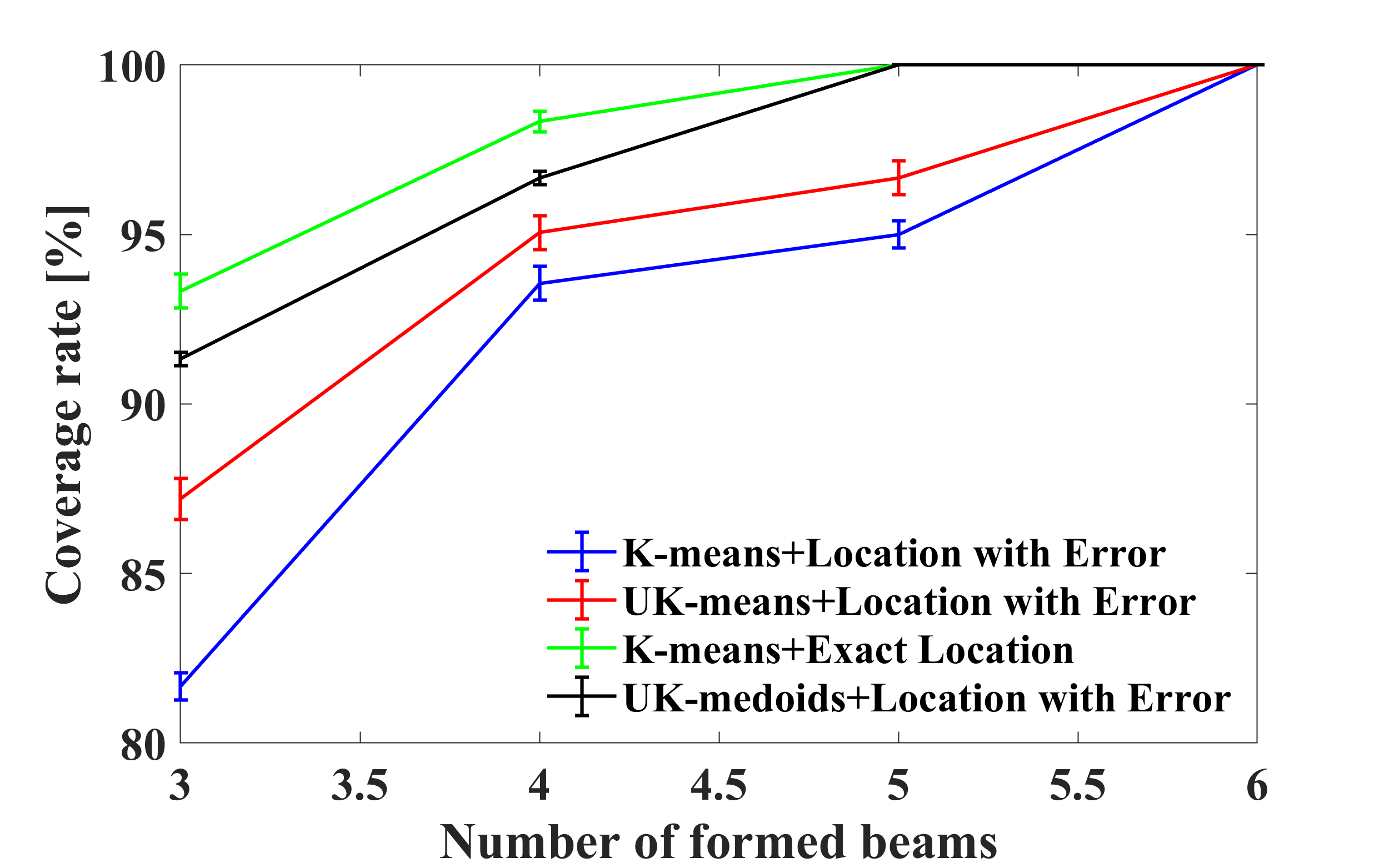}
    \end{minipage}
    }
    \subfigure[Coverage rate under varying beam angle]{
    \begin{minipage}[b]{0.4\textwidth}
    \includegraphics[width=1\textwidth]{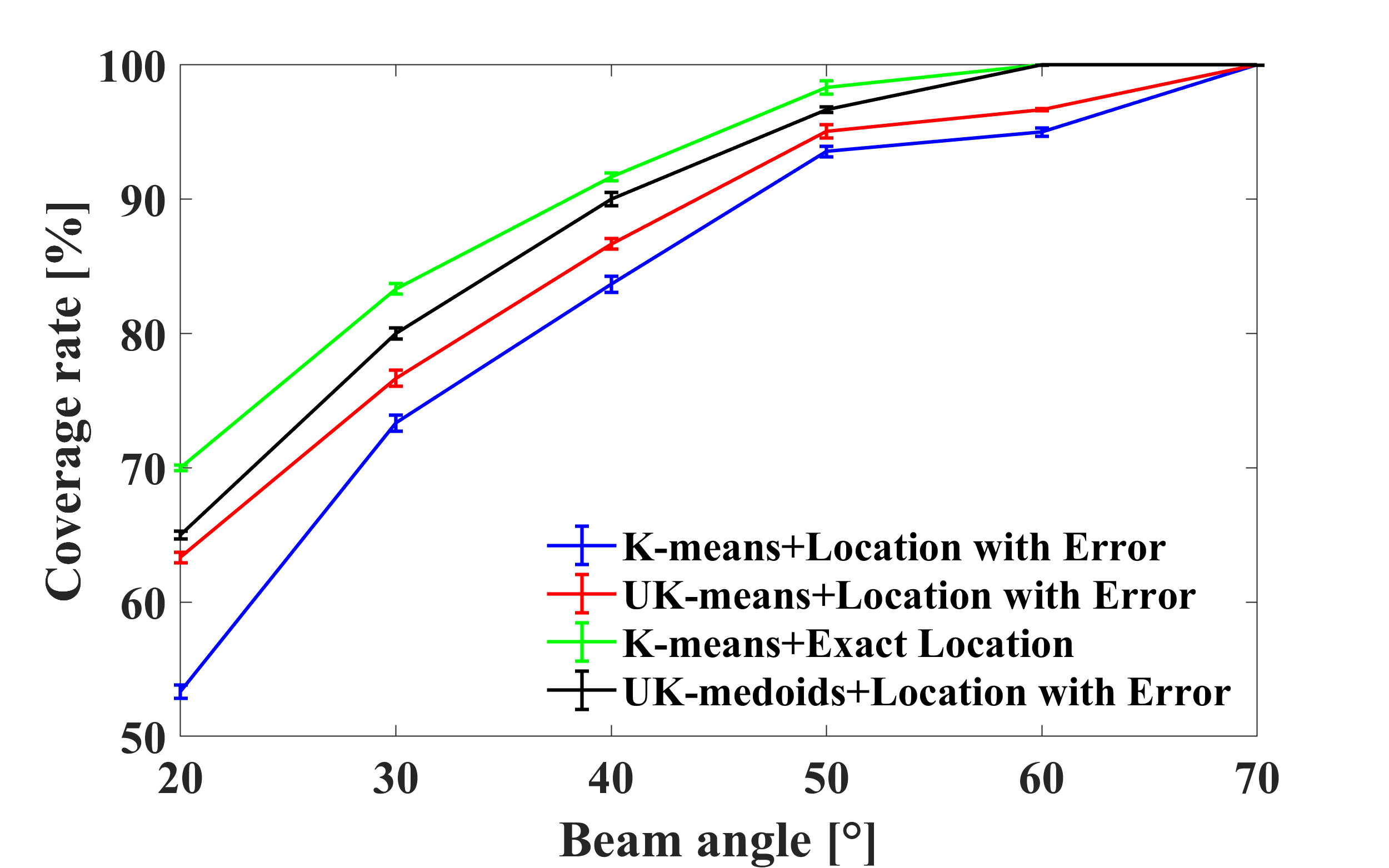}
    \end{minipage}
    }
    \caption{Coverage rate results comparison}
    \label{coverage}
\vspace{-13pt}
\end{figure}

\begin{figure}
    \centering
    \subfigure[Sum rate comparison when error is 17.11m]{
    \begin{minipage}[b]{0.4\textwidth}
    \includegraphics[width=1\textwidth]{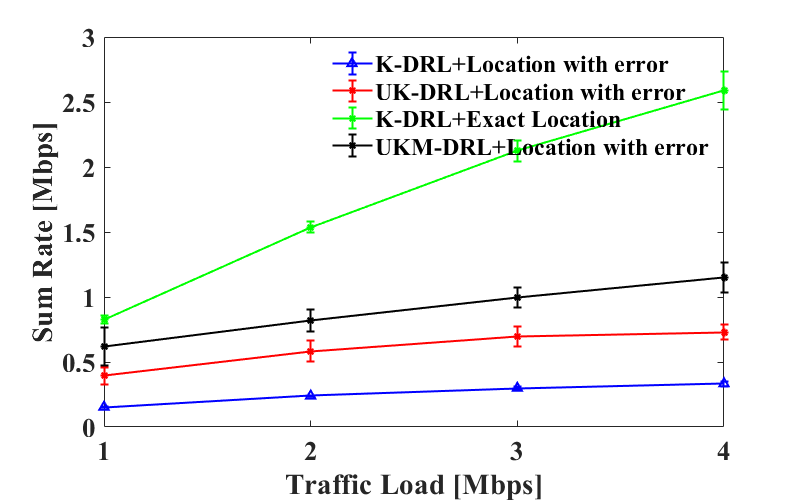}
    \end{minipage}
    }
    \vspace{0pt}
    \subfigure[Sum rate comparison when error is 3.62m]{
    \begin{minipage}[b]{0.4\textwidth}
    \includegraphics[width=1\textwidth]{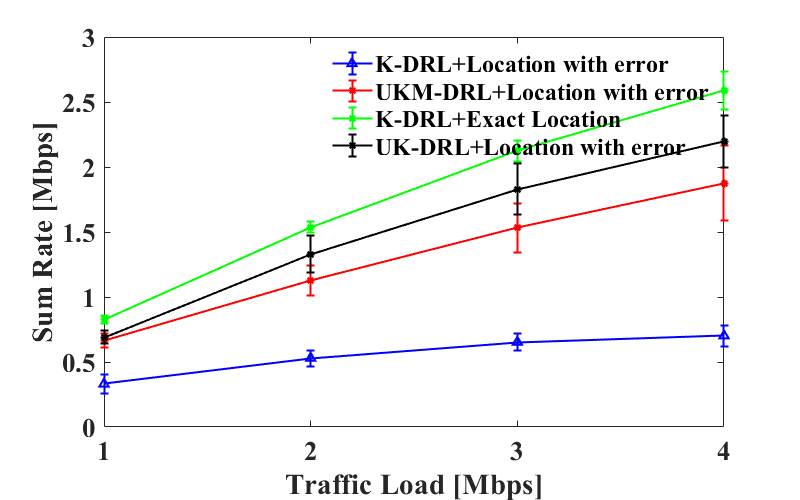}
    \end{minipage}
    }
    \setlength{\abovecaptionskip}{0pt}
    \caption{Sum rate comparison under various traffic loads}
    \label{sumRate}
    \vspace{-15pt}
\end{figure}

\subsubsection{Localization and coverage results}
The localization accuracy is measured by root mean squared error (RMSE) and mean absolute error (MAE). The results of the two measurements are shown in Table \ref{table5}. With the added new features, our proposed method reduces the RMSE to less than 4 meters in each region. For the total localization results, RMSE is reduced from 17.11 meters to 3.62 meters, MAE is reduced from 7.83 meters to 1.51 meters. The main reason is that our vision-aided new features lead to higher accuracy in prediction results.

After clustering, we employ one beam to serve each cluster, then it can be easily found whether one UE is covered by a beam. Then coverage rate is computed by dividing the number of covered UEs by the total number of UEs. Fig. \ref{coverage}(a) shows the coverage rate with different number of beams when beam angle is set to 60$^\circ$. Scenario 3 and 4 need 5 beams to achieve full coverage while scenario 1 and scenario 2 need 6 beams. Since exact location is applied, the coverage rate of scenario 4 is the highest. With the distorted location, UK-medoids shows its strength when compared with UK-means. Meanwhile, K-means has the lowest coverage rate. Fig. \ref{coverage}(b) presents the average coverage rate under different beam angles when the number of beams is 5. It reveals the same conclusion as Fig. \ref{coverage}(a). One can observe that when beam angle is 20$^\circ$, UK-medoids has an improvement of 7.7\% than UK-means.

\subsubsection{Resource Allocation}
In this section, we compare the network performance under different traffic loads and algorithms. As aforementioned, the error is 17.11 meters by using the localization method from \cite{b5}, and the 3.62 meters error is achieved by our proposed method. We present the results under both types of errors to better illustrate the advantage of our proposed UKM-DRL method.

Fig. \ref{sumRate} shows the total data rate of different scenarios under various traffic loads when localization error is 17.11 meters and 3.62 meters. As our optimal baseline, scenario 4 achieves the highest throughput, and the reason is that exact locations allow beams to cover most UEs without any uncertainty. Under both localization errors, UK-medoids outperforms UK-means. K-means produces the worst outcome due to the lack of ability to deal with uncertainty. Moreover, scenarios 1 to 3 can all benefit from reducing the localization error, which is indicated by a higher sum rate. It further demonstrates the capability of our proposed vision-aided localization method.

\begin{figure}
    \centering
    \subfigure[Delay comparison when error is 17.11m]{
    \begin{minipage}[b]{0.4\textwidth}
    \includegraphics[width=1\textwidth]{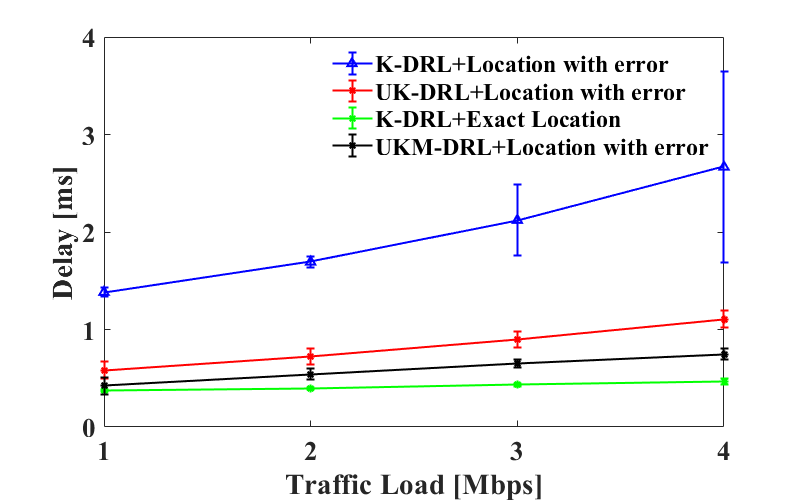}
    \end{minipage}
    }
    \vspace{0pt}
    \subfigure[Delay comparison when error is 3.62m]{
    \begin{minipage}[b]{0.4\textwidth}
    \includegraphics[width=1\textwidth]{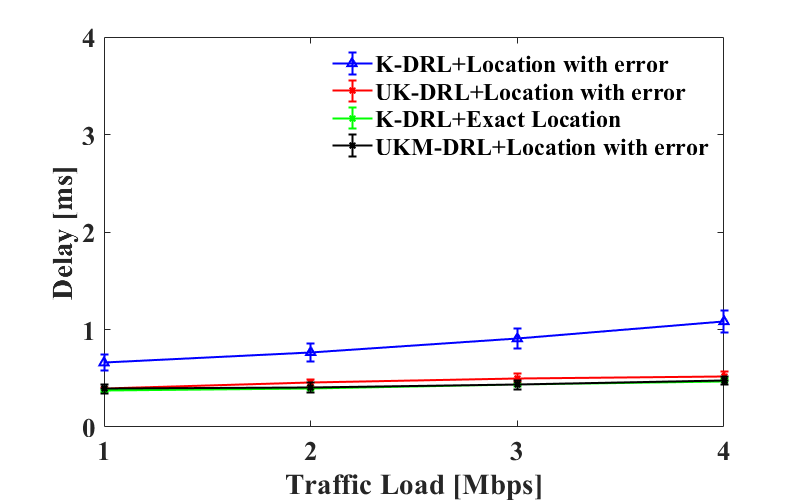}
    \end{minipage}
    }
    \setlength{\abovecaptionskip}{0pt}
    \caption{Delay comparison under various traffic loads}
    \label{delay}
    \vspace{-10pt}
\end{figure}

Similarly, Fig. \ref{delay} presents the delay under different traffic loads when localization error is 17.11 meters and 3.62 meters. The result is consistent with Fig. \ref{sumRate}. 
The optimal baseline presents the lowest latency. With distorted location, UKM-DRL has lower latency than UK-DRL. At last, since K-means has no ability to handle localization error, it has the highest latency. When the error is 3.62 meters, the latency of both UK-DRL and UKM-DRL are very close to the optical baseline. Compared with the results in Fig. \ref{delay}(a), the latency of scenarios 1 to 3 are greatly reduced. When the traffic load is 4 Mbps, UKM-DRL has an improvement of 32.4\% over UK-DRL under 17.11 meters error and 7.7\% under 3.62 meters error.

\section{Conclusion}
\label{s7}
Machine learning is considered as a promising technique to solve the pressing challenges in future 6G networks. In this paper, we investigate the joint vision-aided sensing and communications for beam management of 6G Networks. We first introduce a novel vision-aided localization method to improve localization accuracy. Then we propose a UK-medoids based clustering method to handle the localization uncertainty, and apply the deep reinforcement learning for intra-beam resource allocation. The simulations demonstrate that our proposed method can greatly reduce the localization error, and it achieves lower delay and higher data rate than other baseline algorithms. In the future, we will include more vision-based methods for localization.

\section*{Acknowledgment}
This work is supported by Ontario  Centers  of  Excellence(OCE) 5G ENCQOR program and Ciena, the NSERC Collaborative Research and Training Experience Program (CREATE) under Grant 497981, and Canada Research Chairs Program. We would like to thank Medhat Elsayed and Hind Mukhtar for their help in the earlier versions.

\end{document}